\documentclass[times,twocolumn,final]{elsarticle}

\usepackage{medima}
\usepackage{amsmath,amssymb,amsfonts}
\usepackage{algorithmic}
\usepackage{graphicx}
\usepackage{array}
\usepackage{textcomp}
\usepackage{latexsym}
\usepackage{makecell}
\usepackage{framed,multirow,color}
\usepackage{amsmath}
\usepackage{CJK}
\usepackage{threeparttable}
\usepackage{url}
\usepackage{xcolor}

\usepackage{hyperref}

\definecolor{newcolor}{rgb}{.8,.349,.1}

\journal{Medical Image Analysis}

\begin{document}

\verso{Heng Li \textit{et~al.}}

\begin{frontmatter}

\title{A Generic Fundus Image Enhancement Network Boosted by Frequency Self-supervised Representation Learning\tnoteref{tnote1}}%
\tnotetext[tnote1]{Code of this study is available at https://github.com/liamheng/Annotation-free-Fundus-Image-Enhancement.}

\author[1]{Heng \snm{Li}\fnref{fn1}}
\author[1]{Haofeng \snm{Liu}\fnref{fn1}}
\fntext[fn1]{Contributed equally to this manuscript. e-mail: lih3@sustech.edu.cn, liuhf2020@mail.sustech.edu.cn}
\author[2]{Huazhu \snm{Fu}}
\author[3,4]{Yanwu \snm{Xu}}
\author[5]{Hai \snm{Shu}}
\author[6]{Ke \snm{Niu}}
\author[7]{Yan \snm{Hu}\corref{cor1}}
\author[1,7,8]{Jiang \snm{Liu}\corref{cor1}}
\cortext[cor1]{Corresponding author: 
  huy3, liuj@sustech.edu.cn}
\address[1]{Research Institute of Trustworthy Autonomous Systems and Department of Computer Science and Engineering, Southern University of Science and Technology, Shenzhen, China}
\address[2]{Institute of High Performance Computing (IHPC), Agency for Science, Technology and Research (A*STAR), Singapore}
\address[3]{School of Future Technology, South China University of Technology, Guangzhou, China}
\address[4]{Pazhou Lab,Guangzhou, China}
\address[5]{Department of Biostatistics, School of Global Public Health, New York University, New York, NY, USA}
\address[6]{Computer School, Beijing Information Science and Technology University, Beijing, China}
\address[7]{Department of Computer Science and Engineering, Southern University of Science and Technology, Shenzhen, China}
\address[8]{Guangdong Provincial Key Laboratory of Brain-inspired Intelligent Computation, Southern University of Science and Technology, Shenzhen, China}

\received{2022}
\finalform{2022}
\accepted{2022}
\availableonline{2022}
\communicated{S. Sarkar}

\begin{abstract}
Fundus photography is prone to suffer from image quality degradation that impacts clinical examination performed by ophthalmologists or intelligent systems.
Though enhancement algorithms have been developed to promote fundus observation on degraded images, high data demands and limited applicability hinder their clinical deployment.
To circumvent this bottleneck, a generic fundus image enhancement network (GFE-Net) is developed in this study to robustly correct unknown fundus images without supervised or extra data.
Levering image frequency information, self-supervised representation learning is conducted to learn robust structure-aware representations from degraded images. 
Then with a seamless architecture that couples representation learning and image enhancement, GFE-Net can accurately correct fundus images and meanwhile preserve retinal structures.
Comprehensive experiments are implemented to demonstrate the effectiveness and advantages of GFE-Net.
Compared with state-of-the-art algorithms, GFE-Net achieves superior performance in data dependency, enhancement performance, deployment efficiency, and scale generalizability.
Follow-up fundus image analysis is also facilitated by GFE-Net, whose modules are respectively verified to be effective for image enhancement.
\end{abstract}


\end{frontmatter}



\section{Introduction}
\label{sec:introduction}
\begin{figure}[tp]
\centering
\includegraphics[width=0.9\linewidth]{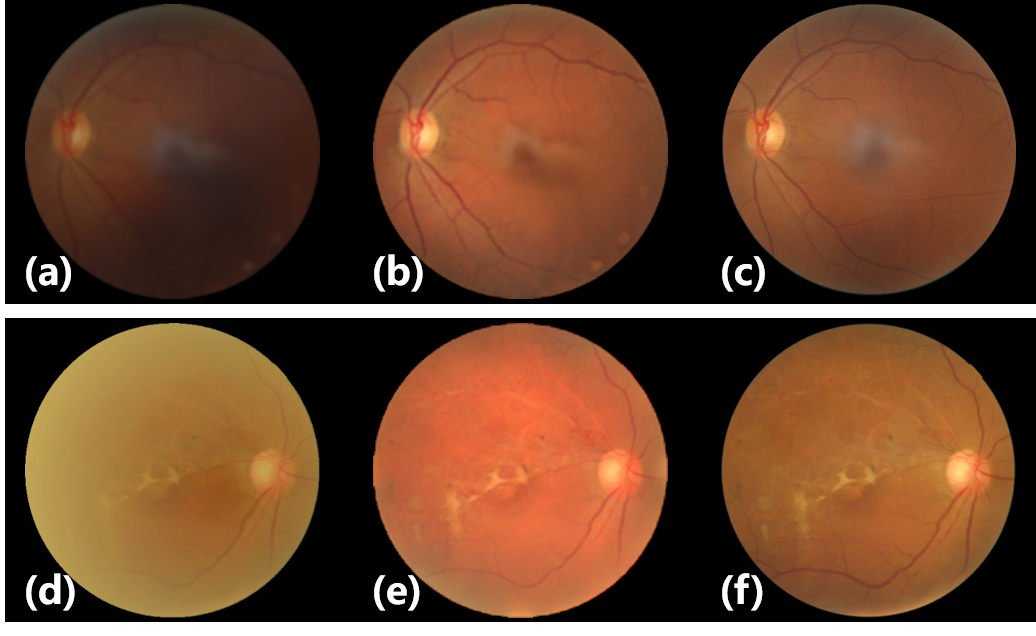}
\caption{
Illustration of fundus image enhancement. (a) and (d) exhibit fundus images respectively with imaging and cataract interferences. (b) and (e) are the enhancement results of the proposed algorithm. (c) and (f) visualize the high-quality references.
}
\label{fig:example}
\end{figure}

Owing to the superiority in safety and cost, fundus photography has been used as a routine clinical examination to diagnose and monitor ocular diseases~\citep{li2021applications}, such as diabetic retinopathy (DR), glaucoma, and age-related macular degeneration (AMD). 
Unfortunately, fundus images are prone to quality degradation due to imaging or cataract interferences~\citep{macgillivray2015suitability}, leading to uncertainties in clinical observations.
According to screening studies~\citep{philip2005impact,teng2002progress}, over 10\% of mydriatic and 20.8\% of non-mydriatic fundus images are of a quality that is unreadable to ophthalmologists. 
In UK Biobank, only 71.5\% of 135,867 fundus images meet the quality standards required for vessel morphometry~\citep{welikala2017automated}.
On the other hand, cataracts, by attenuating and scattering the light passing through the lens, result in degraded fundus images, which cannot be solved with repeated photography for cataract patients~\citep{flaxman2017global}.
Fig.~\ref{fig:example} exhibits fundus images affected by interferences, where the top row illustrates the impact from imaging interferences, and the bottom row is an image from a cataract patient.
Compared with the high-quality references in Fig.~\ref{fig:example} (c) and (f), it is intractable to clearly observe fundus details in the degraded images in Fig.~\ref{fig:example} (a) and (d).
Therefore, unreadable image quality not only prevents reliable diagnosis by ophthalmologists, but also impacts the performance of intelligent fundus assessment systems.

An intuitive solution to improve the certainty in fundus observation is enhancing the readability of fundus images (see Fig.~\ref{fig:example} (b) and (e)), and efforts have hence been made to develop enhancement algorithms~\citep{shen2022image}. 
Histogram equalization~\citep{setiawan2013color}, spatial filtering~\citep{cheng2018structure}, and frequency filtering~\citep{cao2020retinal,mitra2018enhancement} were leveraged to correct fundus images. 
Whereas these hand-crafted methods are based on prior knowledge of statistics, resulting in insensitivity to retinal details and instability to clinical variations. 
In recent years, the strong image embedding capability of deep learning has been introduced to adaptively enhance fundus images. 
Training and data augmentation techniques have been subsequently developed to mitigate the data dependency of enhancement models~\citep{shen2020modeling,ma2021structure,li2022structure,liu2022degradation}. 
However, unpaired or synthesized data result in suboptimal solutions, unfavorable for structure-preservation and model stability. 
More recently, to properly correct real fundus images based on synthesized data, domain adaptation has been used to generalize the model to real data~\citep{li2022annotation} by visiting test data during the training phase.

Collecting sufficient high-low quality corresponding data is extremely expensive and even unreachable in clinics.
Though solutions have been reported in pioneer image enhancement studies to alleviate the data dependency, clinical challenges remain in algorithm deployment:
i) Though data dependency has been alleviated by previous studies, unpaired data and access to test data are still commonly necessary.
Therefore, extra data collecting and repeated model training are inevitable to deploy the enhancement algorithms in clinics;
ii) To preserve retinal structures in the enhancement, structure guidances, such as importance maps and segmentation outcomes, have been introduced to highlight retinal structure details.
Unfortunately, the guidances are also sensitive to image degradations, leading to unreliable structural clues;
iii) While adversarial learning promotes the performance of state-of-the-art (SOTA) algorithms, it is liable to result in a time-consuming training phase and limited generalizability to unknown data.

Considering the above challenges, a generic fundus image enhancement network, termed GFE-Net, is developed in this paper to robustly enhance unknown low-quality fundus images only based on synthesized data (as shown in Fig.~\ref{fig:example}).
A self-supervised representation learning (SSRL) strategy is proposed to learn robust structure-aware representations using frequency information of fundus images.
Subsequently, representation learning and image enhancement are seamlessly coupled by GFE-Net to properly correct unknown fundus images without visiting any supervised or test data.
The main contributions of this paper are summarized as follows:
\begin{itemize}
  \item [1)] To conveniently deploy fundus image enhancement in clinics, GFE-Net is proposed with a seamless SSRL architecture to robustly correct unknown fundus images only based on synthesized data.
  \item [2)] Frequency self-supervision is designed and integrated in the enhancement model optimization to capture robust structure-aware representations, thereby preserving fine structures when correcting fundus images with unknown degradations.
  \item [3)] The elegant seamless SSRL architecture, coupled with explicit objective functions, endows GFE-Net with distinct advantages for clinical deployments, including the efficient achievement of training convergence and the elimination of additional tuning with clinical data.
  \item [4)] Extensive experiments are conducted to validate the benefits of GFE-Net. Compared with SOTA algorithms, the specially designed modules within GFE-Net promise superior performances in fundus image enhancement with alleviated dependency and strong generalizability.
\end{itemize}


\section{Related Work}
\subsection{Hand-crafted fundus image enhancement}
Hand-crafted algorithms have been tailored to enhance fundus images in previous studies.
Contrast enhancement, which expands the dynamic ranges of images, is an effective method for improving image readability.
Therefore, contrast limited adaptive histogram equalization (CLAHE)~\citep{zuiderveld1994contrast} has been introduced to restore degraded fundus images~\citep{mitra2018enhancement,setiawan2013color}.
Then frequency filters have also been combined with CLAHE to enhance the fundus image contrast~\citep{mitra2018enhancement}.
Moreover, low-pass filters and $\alpha$-rooting have been implemented to improve the contrast of the retinal structures~\citep{cao2020retinal}. 
Alternatively, spatial filters have been developed to estimate and remove the noise in imaging models. Dark channel prior~\citep{he2010single} and guided image filtering (GIF)~\citep{he2012guided} have been extensively employed in image enhancement and restoration. 
Based on GIF, structure-preserving guided retinal image filtering (SGRIF)~\citep{cheng2018structure} has been designed to promote structure preservation when restoring cataract-affected fundus images.
However, indiscriminately processing the entire image leads hand-crafted algorithms to either struggle to preserve fine details or suffer from strict constraints in implementation.

\subsection{Deep learning based fundus image enhancement}
In recent years, thanks to the superiority in image representation, deep learning has been frequently applied in computer vision, accelerating the advancements in image restoration~\citep{lore2017llnet,chen2018robust,huang2021neighbor2neighbor}.
With a number of fully supervised training data, deep learning allows enhancement approaches to adaptively learn the mapping from a low-quality image to the corresponding high-quality one~\citep{isola2017image,deng2022rformer}.
Unfortunately, collecting an abundant amount of supervised data is often impractical in medical scenarios. 
As a result, recent algorithms strive to learn the mapping from unpaired or synthesized data.

\subsubsection{Training with unpaired data}
Based on CycleGAN~\citep{zhu2017unpaired}, StillGAN~\citep{ma2021structure} was proposed to learn a suitable mapping from a low-quality domain to a high-quality domain, or from the domain with non-uniform illumination to a high-quality one.
Besides, inspired by contrastive unpaired translation (CUT)~\citep{park2020contrastive}, I-SECRET~\citep{cheng2021secret} was developed to enhance fundus images via contrastive learning.
Nevertheless, learning with unpaired data tends to be less effective in preserving retinal structures, resulting in suboptimal enhancement performance.

\subsubsection{Training with synthesized data}
Alternatively, high-low quality paired data were synthesized to enforce structure preservation in fundus image enhancement. 
Cataract fundus images were generated via image-to-image translation~\citep{luo2020dehaze} to train enhancement models. 
Through modeling the imaging interferences in fundus photography, fundus images were deliberately degraded to train a clinical-oriented fundus enhancement network (CofeNet)~\citep{shen2020modeling} under the supervision of the original ones. 
Furthermore, domain adaptation~\citep{li2021Restoration} was introduced to bridge the domain shift between the synthesized and real data.
{
MAGE-Net~\citep{guo2023bridging} utilizes a transferable and multiple consistency approach to bridge the gap between synthetic and real fundus image enhancement.
}
In our previous study~\citep{li2022annotation}, test data were accessed during training to generalize the model learned from synthesized cataract data to real ones by unsupervised domain adaptation.

Although recent algorithms have shown performance improvements in fundus image enhancement, their reliance on additional annotations or data poses limitations on their clinical deployment.

\subsection{Self-supervised representation learning}
Representation learning drives machine learning algorithms to discover effective representations for the downstream tasks~\citep{bengio2013representation}.
Supervised representation learning extracts representations to solve other related tasks based on annotated data from a specific task. 
However, the supervision relies on plenty of annotated data, and may not provide the ideal representations for the tasks suffering from domain shifts.
To alleviate the annotation and data bottleneck in practical deployments, SSRL~\citep{ericsson2022self} has been proposed to provide a powerful framework for deep feature learning without the dependency on large annotated data sets.

By carefully designing pretext tasks, SSRL produces freely available labels that can serve as supervision to learn deep representations and can be reused to solve downstream tasks with comparatively small task-specific annotated data. Transformation prediction and instance discrimination are two major self-supervised pretexts.
Transformation prediction consists of several methods, such as coloring~\citep{larsson2016learning} or inpainting~\citep{pathak2016context} masked-out information in images, and predicting the rotation angle~\citep{komodakis2018unsupervised} of images.
To correctly solve the pretext task, the information regarding the transformation needs to be retained in the representation.
Self-supervised contrastive learning,
as demonstrated in recent studies~\citep{bachman2019learning,chen2020simple,grill2020bootstrap}, has emerged as a remarkable paradigm for instance discrimination, whose core idea is to pull together positive pairs and push apart negative pairs.
Nevertheless, contrastive learning preserves the shared information between positive pairs and eliminates the non-shared information~\citep{wang2022rethinking}. As a result, it cannot be guaranteed that all task-relevant features are contained in the learned representations~\citep{tian2020makes}.

To boost the clinical deployment of fundus image enhancement, 
in this paper,
an SSRL paradigm using frequency information is designed and seamlessly coupled with the down-steam task to robustly correct fundus images without extra annotations or data training dependency.

\section{Method}
\label{sec:method}
To efficiently and steadily improve the readability of clinical low-quality fundus images, a generic fundus enhancement network is developed to present robust enhancement without extra annotations or data.
Using the frequency self-supervision and synthesized high-low quality image pairs, representation learning and image enhancement are seamlessly cooperated to robustly correct fundus images and preserve retinal structures. 

\subsection{Frequency self-supervised representation learning}
Given that retinal structures are the foundation of fundus assessment, 
it is essential to preserve the structures in fundus image enhancement.
Inspired by the self-supervised pretexts of transformation prediction and contrastive learning, 
we employ
the frequency attributes of fundus images to perform SSRL to learn robust representations of retinal structures.

\subsubsection{Degraded view dataset}
\label{sec:CDC}
Collecting augmented views of a sample is a commonly adopted strategy to learn effective representations in contrastive learning~\citep{chen2020simple,wang2022rethinking}. 
Motivated by this, we synthesize various degraded views from clear fundus images to learn representations robust to image degradations.

Specifically, from an individual clear image, a set of degraded views is randomly synthesized, such that variant degradations and consistent retinal structures are included, which can be leveraged to learn robust representations.
As reported in~\cite{shen2020modeling} and~\cite{li2022annotation}, imaging interferences and cataracts are the major degradation factors of fundus images, and style randomization of fundus images was proposed in~\cite{liu2021feddg}.
Therefore, degraded views can be generated using degradation simulation and style randomization.
For a clear image $x\in \mathbb{R}^K $, the degraded view set is given by
\begin{equation}
\{\tilde{x}_{v}\} = \{\delta_{v} (x) \mid v=1,2,...,V\},
\label{eq:views}
\end{equation}
where $\tilde{x}_{v}\in \mathbb{R}^{K\times V}$, $\delta_{v} (\cdot )$ denotes the fundus image degradations modeled following~\cite{shen2020modeling}, \cite{liu2021feddg} and~\cite{li2022annotation}, and $v$ is the index of the view sample which records the models and parameters used in the degradation.

As shown in Fig.~\ref{fig:workflow}, the degraded view dataset is first collected from a clear image with changing simulation models and parameters, such that various degradations and consistent retinal structures are contained in the dataset.

\begin{figure}[bp]
\centering
\includegraphics[width=\linewidth]{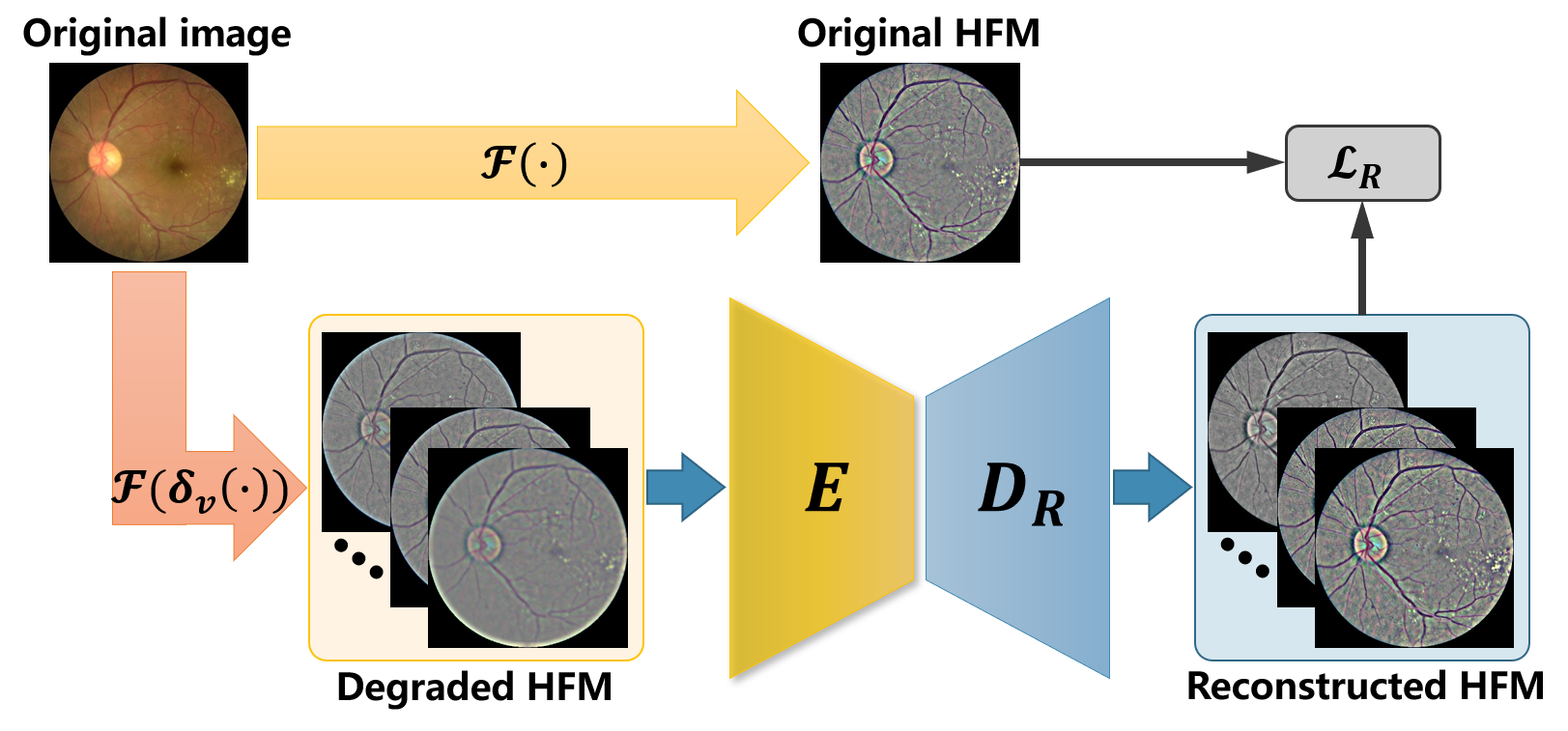}
\caption{SSRL using image frequency information. As retinal structures concentrate on HFM, $E$ and $D_R$ learn robust structure-aware representations from reconstructing HFM from degraded images.
}
\label{fig:fss}
\end{figure}

\begin{figure*}[htbp]
\centering
\includegraphics[width=\linewidth]{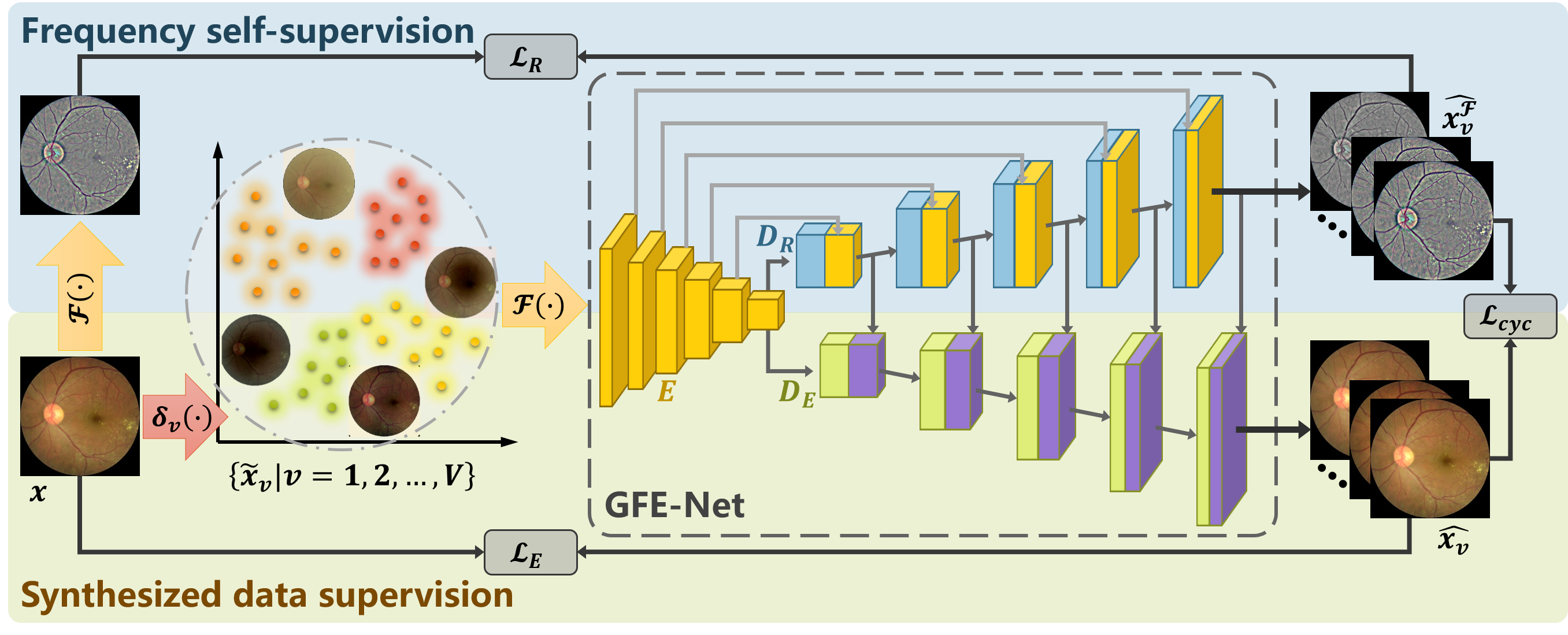}
\caption{
Workflow of the proposed algorithm. 
Frequency self-supervision and synthesized data supervision are cooperated to train GFE-Net.
Based on a clear image $x$, a degraded view dataset $\{\tilde{x}_{v}\}$ is synthesized. 
Then GFE-Net seamlessly couples representation learning and image enhancement, where structure-aware representations are learned in $E$ and $D_R$ by SSRL (shown in Fig.~\ref{fig:fss}) and further forwarded to correct the fundus images in $D_E$.
}
\label{fig:workflow}
\end{figure*}

\subsubsection{Frequency self-supervision}
Properly reconstructing fine structures is crucial for correcting degraded images, especially fundus images.
According to the Retinex theory~\citep{land1977retinex}, retinal structures concentrate in the high-frequency domain of fundus images.
Thus, as visualized in Fig.~\ref{fig:fss}, an SSRL strategy based on frequency information is proposed to learn robust representations of retinal structures, circumventing the need for segmentation~\citep{luo2020dehaze,shen2020modeling}.

As the structure details of a clear image are concentrated in its high-frequency map (HFM)~\citep{li2023frequency}, a pretext task is established to discover robust representations for preserving retinal structures.
Based on the clear image $x$ and degraded views $\{\tilde{x}_{v}\}$, the HFM of $x$ is regarded as a frequency self-supervision to be reconstructed from that of $\{\tilde{x}_{v}\}$.
Through the above pretext task, a reconstruction network is trained to extract structure-aware representations robust to image degradation.
In practice, the pretext task is performed with a high-pass filter $\mathcal{F}(\cdot)$ based on the Gaussian kernel, which is given by

\begin{equation}
\mathcal{F}(x)= x - x\ast g(r,\sigma),
\label{eq:filter}
\end{equation}
where $g$ is a Gaussinan filter with the kernel of $(r,\sigma)$. 

\subsubsection{Representation learning objectives}
With the degraded views and frequency self-supervision, SSRL is conducted by reconstructing the HFM of clear images from degraded ones.
Since structure details are concentrated in the high-frequency domain, robust representations for retinal structures can be learned from the HFM reconstruction.
HFMs are captured respectively from the clear image $x$ and its degraded views, serving as the supervision and inputs for an image reconstruction network.

The reconstruction network is built with a U-Net and the optimization objective is given by
\begin{equation}
\mathcal{L}_R(E,D_R)=\mathbb{E}\left [{\textstyle\sum_{v=1}^{V}}\left \| \mathcal{F}(x)-\widehat{x^{\mathcal{F}}_v} \right \|_{1}\right ],
\label{eq:lh}
\end{equation}
where $\widehat{x^{\mathcal{F}}_v} = D_R(E(\mathcal{F}(\tilde{x}_{v})))$ denotes the reconstructed HFM, $E$ represents the image encoder, and $D_R$ denotes the representation learning decoder.

\subsection{GFE-Net coupled with representation learning}
Through the frequency self-superivsion, robust structure representations can be captured in $E$ and $D_R$. 
Subsequently, the representations need to be suitably introduced for fundus image enhancement.
As demonstrated in Fig.~\ref{fig:workflow}, representation learning and image enhancement are seamlessly coupled to correct fundus images.
Specifically, the GFE-Net is composed of the encoder $E$, decoder $D_R$, and an enhancement decoder $D_E$.
The structure-aware representations are thus fully availed to preserve retinal structures in the enhancement.

\subsubsection{Shared encoder}
The encoder $E$ is shared by $D_R$ (blue blocks) and $D_E$ to conduct both representation learning and image enhancement.
Therefore, $E$ is optimized by the complementary inductive bias between different tasks.
Additionally, skip connections are bridged between $E$ and $D_R$ to form a U-Net to forward low-level features.

\subsubsection{Representation learning decoder}
Supervised by the HFM of clear images, $D_R$ reconstructs the HFM of degraded views to learn robust structure-aware representations.
The input vector of the $l$-th layer of $D_R$ is formulated by
\begin{equation}
f^l_R=[D^{l-1}_R(f^{l-1}_R),f^{L-l+1}],\;l=2,3,...,L,
\label{eq:fh}
\end{equation}
where $D_R^l$ denotes the $l$-th layer of $D_R$, and $f^l$ is the features captured by the $l$-th layer of $E$.
Notably, $f^1_R=f^L$, where $L$ is the total number of layers in $E$.

\subsubsection{Image enhancement decoder}
By learning the structure-aware representations consistent with clear images, robust representations are extracted in $E$ and $D_R$.
Consequently, $E$ shares structure-aware inputs between $D_R$ and $D_E$, and then vectors from $D_R$ further boost structure preservation in the enhancement by $D_E$.

$D_E$ is coupled with $D_R$ using the same architecture to import the representations in $D_R$.
As demonstrated in Fig.~\ref{fig:workflow}, the outputs from one layer of $D_R$ and $D_E$ are concentrated as the input to the next one of $D_E$.
The input vector of the $l$-th layer of $D_E$ is given by
\begin{equation}
f^l_E=[D^{l-1}_E(f^{l-1}_E),D^{l-1}_R(f^{l-1}_R)],\;l=2,3,...,L,
\label{eq:fr}
\end{equation}
where $D_E^l$ is the $l$-th layer of $D_E$, and $f^1_E=f^L$.

\subsubsection{Image enhancement objectives}
The enhancement by GFE-Net is supervised by the synthesized high-low quality image pairs.
With the enhancement outcomes of $\tilde{x}_{v}$ denoted as $\widehat{x_v}$, an enhancement loss is calculated by
\begin{equation}
\mathcal{L}_{E}(E,D_R,D_E)=\mathbb{E}\left[{\textstyle\sum_{v=1}^{V}}\left \| x-\widehat{x_v} \right \|_{1}\right].
\label{eq:lr}
\end{equation}

Moreover, if $D_R$ desirably reconstructs the HFM and $D_E$ properly enhances fundus images, the HFM of $\widehat{x_v}$ should be consistent with $\widehat{x^{\mathcal{F}}_v}$.
Therefore, a cycle-consistency loss is computed by
\begin{equation}
\mathcal{L}_{cyc}(E,D_R,D_E)=\mathbb{E}\left[{\textstyle\sum_{v=1}^{V}}\left \| \mathcal{F}(\widehat{x_v}) - \widehat{x^{\mathcal{F}}_v} \right \|_{1}\right].
\label{eq:lcyc}
\end{equation}

Finally, cooperating with the objective of representation learning, the total objective function for tuning GFE-Net is defined as
\begin{equation}
\begin{aligned}
\mathcal{L}_{total}(E,D_R,D_E)=&\mathcal{L}_R(E,D_R)+\mathcal{L}_{E}(E,D_R,D_E)\\
&+\mathcal{L}_{cyc}(E,D_R,D_E).
\end{aligned}
\label{eq:overall}
\end{equation}
Notably, instead of the pretrained and fine-tuned framework, representation learning and image enhancement are seamlessly coupled in the architecture of GFE-Net.
Therefore, convergence can be efficiently achieved in training with Eq.~\ref{eq:overall}, where model optimization is only based on explicit objective functions, circumventing model collapse and vanishing gradients in adversarial learning.

\subsection{Network implementation}
For GFE-Net, $E$ consists of 8 down-sampling layers, whose down-sampling layer contains a down convolution layer, a leaky ReLU layer, and a batch normalization layer. 
$D_R$ and $D_E$ consist of 8 up-sampling layers, whose up-sampling layer contains a transposed convolution layer, a ReLU layer, and a batch normalization layer, while an output layer contains a ReLU layer, a convolution layer, and a Tanh function. The convolutional and deconvolutional layers have a kernel size of 4 $\times$ 4 and a stride of 2.

In the training phase, the input image size was 256 $\times$ 256 and the batch size was 8. 
The training data were loaded with a random scale among 286, 306, 326, and 346, and then cropped to the size of 256 $\times$ 256.
The model was trained by the Adam optimizer for 150 epochs with an initial learning rate of 0.001 and 50 epochs with the learning rate gradually decaying to 0. 
{
Comparative algorithms were implemented using public code with the same learning rates and number of epochs.
}

\begin{table*}[tbp]
\footnotesize
\centering {\caption{Datasets and evaluation metrics used in the experiments}
\label{tab:dataset} }%
\renewcommand{\arraystretch}{1.1}
\setlength\tabcolsep{6pt}
\begin{tabular}{m{1.2cm}| m{3.5cm}  m{2.3cm}|  m{4.2cm}   m{4.2cm} }
\hline
\multirow{2}{1.6cm}{Evaluation} & \multicolumn{2}{c|}{With reference} & \multicolumn{2}{c}{Without reference} \\
\cline{2-3} \cline{4-5}
& Enhancement & Segmentation & Enhancement & Diagnosis \\
\hline
Metrics & SSIM, PSNR & IoU, Dice & FIQA, WFQA & F1-score, Ckappa\\
\hline
\makecell[*l]{Training} & \makecell[*l]{$\nabla$ DRIVE: 40 clear images} && \makecell[*l]{$\nabla$ DRIVE: 40 clear images} & \makecell[*l]{$\nabla$ 7,331 clear images in Fundus-iSee}\\
\hline
\makecell[*l]{Test} & \makecell[*l]{$\triangle$ FIQ: 196 low-high quality fundus image pairs, \\ $\triangle$ RCF: 26 pre- and post-operative image pairs} && \makecell[*l]{$\triangle$ EyeQ-Usable, Reject, \\ $\triangle$ 100 cataract images from Kaggle} & $\triangle$ 2,669 cataract images in Fundus-iSee\\
\hline
\end{tabular}
\end{table*}

\section{Experiments}
Experiments were implemented to interpret the performance of the proposed algorithm. 
Data requirements, enhancement performance, deployment efficiency, and scale generalizability were comprehensively compared with SOTA algorithms to demonstrate the superiority of GFE-Net.
Then GFE-Net was deployed to boost the segmentation and diagnosis of degraded fundus images.
Subsequently, the effectiveness of training data and designed modules were presented in the ablation study. 

\subsection{Experiment Setting}
In the experiments, generic enhancement models were constructed and verified with six fundus datasets.
To fully illuminate the performance of GFE-Net, eight evaluation metrics of fundus image enhancement, segmentation, and diagnosis were calculated, and nine SOTA algorithms were compared as baselines.
The experimental protocol is in accordance with the Declaration of Helsinki and was approved by the local Ethics Committee.

\subsubsection{Datasets}
The experiments used six datasets:

\noindent $\triangleright$ \textbf{DRIVE}\footnote[1]{http://www.isi.uu.nl/Research/Databases/DRIVE/}: 40 clear fundus images with segmentation masks.

\noindent $\triangleright$ \textbf{FIQ}: a private dataset acquired from Shenzhen Kangning Hospital, contains 196 low-high quality image pairs acquired from repeated fundus examination.

\noindent $\triangleright$ \textbf{RCF}: a private dataset collected by Peking University Third Hospital consists of 26 fundus images after cataract surgery corresponding to the ones before surgery.

\noindent $\triangleright$ \textbf{EyeQ}~\citep{fu2019evaluation}: a subset of EyePACS\footnote[2]{https://www.kaggle.com/c/diabetic-retinopathy-detection}, contains 28,792 samples (16,817 `Good', 6,435 `Usable', and 5,540 `Reject') sorted according to image quality.

\noindent $\triangleright$ \textbf{Kaggle}\footnote[3]{https://www.kaggle.com/jr2ngb/cataractdataset}: 300 clear fundus images and 100 cataract ones.

\noindent $\triangleright$ \textbf{Fundus-iSee}: a private fundus dataset including 10,000 images (2,669 with and 7,331 without cataracts), annotated into five categories according to fundus status.

As summarized in Table~\ref{tab:dataset}, paired data were synthesized from DRIVE to train GFE-Net, and other datasets were also visited according to the data dependency of SOTA algorithms.
FIQ and EyeQ were degraded by imaging interferences, whereas RCF, Kaggle, and Fundus-iSee suffer from cataracts.
Full reference evaluations of enhancement and segmentation were carried out on FIQ and RCF, while non-reference enhancement evaluations were conducted on EyeQ and Kaggle. 
Additionally, to adapt automatic diagnosis, both diagnosis and enhancement models were learned from clear images in Fundus-iSee.

\begin{table*}[tbp]
\footnotesize
\centering {\caption{{Comparisons in data dependency, enhancement performance, and deployment efficiency.}}
\label{tab:comparisons} }%
\renewcommand{\arraystretch}{1.1}
\begin{threeparttable}
\begin{tabular}{p{37mm}| p{1.5mm}<{\centering} p{1.5mm}<{\centering} p{1.5mm}<{\centering} p{3mm}<{\centering}| p{4mm}<{\centering} p{4mm}<{\centering} p{0.0mm} p{4mm}<{\centering} p{4mm}<{\centering} p{0.0mm} p{4mm}<{\centering} p{4mm}<{\centering} p{0.0mm} p{4mm}<{\centering} p{6.5mm}<{\centering}| p{7mm}<{\centering} p{7mm}<{\centering} p{8mm}<{\centering}}
\hline
\multirow{3}{*}{Algorithms} & \multicolumn{4}{c|}{Data dependency*} & \multicolumn{11}{c|}{Enhancement performance} & \multicolumn{3}{c}{Deployment efficiency}\\
\cline{2-19}
& \multirow{2}{*}{\textsl{HF}} & \multirow{2}{*}{\textsl{SM}} & \multirow{2}{*}{\textsl{UP}} & \multirow{2}{*}{\textsl{TD}} & \multicolumn{2}{c}{FIQ} && \multicolumn{2}{c}{EyeQ} && \multicolumn{2}{c}{RCF} && \multicolumn{2}{c|}{Kaggle} & Costs & Training & Inference\\
\cline{6-7}
\cline{9-10}
\cline{12-13}
\cline{15-16}
& & & & & SSIM & PSNR && FIQA & WFQA && SSIM & PSNR && FIQA & WFQA & (GMac) & (Hour) & (Second)\\
\hline
SGRIF~\citep{cheng2018structure}& \textcolor{red}{$\star$} & & & & 0.776 & 19.20 && 0.18 & 0.55 && 0.609  & 15.07 && 0.17 & 0.70 & 0.03 & --  & 0.13\\
RFormer~\citep{deng2022rformer}&  & & & & 0.788 & 16.61 && 0.44 & 1.14 && 0.728 & 17.14 && 0.59 & 1.29  & 45.46 & -- & 0.16\\
CycleGAN~\citep{zhu2017unpaired}& & & \textcolor{red}{$\star$} & & 0.848 & 20.59 && 0.53 & 1.23 && 0.725 & 17.55 && 0.56 & 1.38 & 56.89 & 23.84 & 0.18\\
I-SECRET~\citep{cheng2021secret} & & & \textcolor{red}{$\star$} & \textcolor{red}{$\star$} & 0.868 & 21.32 && 0.55 & 1.38 && 0.750 & 18.49 && 0.50 & 1.31 & 56.88 & 16.88 & 0.18\\
StillGAN~\citep{ma2021structure} & & & \textcolor{red}{$\star$} & & 0.871 & 21.44 && 0.73 & 1.54 && 0.748 & 18.24 && 0.54 & 1.29 & 67.12 & 51.71 & 0.20\\
\cite{luo2020dehaze}& & \textcolor{red}{$\star$} & & \textcolor{red}{$\star$} & 0.827 & 16.91 && 0.65 & 1.46 && 0.742 & 17.71 && 0.46 & 1.06 & 40.12 & 9.47 & 0.14\\
CofeNet~\citep{shen2020modeling} & & \textcolor{red}{$\star$} & & & 0.838 & 20.64 && 0.76 & 1.67 && 0.744 & 17.83 && 0.48 & 1.29 & 67.50 & 19.76 & 0.19\\
MAGE-Net~\citep{guo2023bridging} & & \textcolor{red}{$\star$} & & \textcolor{red}{$\star$} & 0.861 & 21.64 && 0.52 & 1.35 && 0.762 & 18.12 && 0.69 & 1.61 & 854.92 & 16.23 & 0.27\\
ArcNet~\citep{li2022annotation} & & & & \textcolor{red}{$\star$} & 0.868 & 21.51 && 0.78 & 1.63 && 0.760 & 18.36 && 0.77 & 1.63 & \textbf{18.16} & 6.84 & \textbf{0.10}\\
GFE-Net (ours)& & & & & \textbf{0.879} & \textbf{22.19} && \textbf{0.81} & \textbf{1.65} && \textbf{0.771} & \textbf{18.59} && \textbf{0.78} & \textbf{1.68} & 34.80 & \textbf{3.43} & 0.14\\ %
\hline
\end{tabular}
\begin{tablenotes}
 \scriptsize
 \item{*} Data dependency on hand-crafted features (\textsl{HF}), segmentation masks (\textsl{SM}), unpaired training data (\textsl{UP}), and access to test data (\textsl{TD}) are indicated by \textcolor{red}{$\star$}.
\end{tablenotes}
\end{threeparttable}
\end{table*}

\subsubsection{Evaluation metrics}
For datasets with reference images, the 
structural similarity (SSIM) and the peak signal to noise ratio (PSNR) were used to quantify the enhancement performance.
The segmentation accuracy was evaluated by the intersection over union (IoU) and the Dice coefficient between the restored images and the reference.

For datasets without reference images, MCF-Net~\citep{fu2019evaluation} was employed to quantify the enhancement performance of fundus images.
A fundus image quality assessment score (FIQA) is computed according to the ratio of images identified as `Good' quality by MCF-Net~\citep{cheng2021secret}.
Moreover, a weighted FIQA (WFQA)~\citep{liu2022degradation} is introduced to delicately evaluate the fundus image quality, which is calculated by assigning the weights of 2, 1 and 0 to the images respectively identified as `Good', `Usable' and `Reject' and then dividing by the total number of images.

The automatic diagnosis of fundus diseases was evaluated by the two metrics, F1-score and Cohen's kappa (Ckappa)~\citep{jeni2013facing}.

\subsubsection{Baselines}
\label{baseline}
To demonstrate the benefits of GFE-Net, a comparative analysis was conducted against existing methods in the experiments. Nine SOTA algorithms for fundus image enhancement or restoration were implemented as baselines. 

Based on spatial filtering, SGRIF~\citep{cheng2018structure} was developed to restore cataract fundus images.
CycleGAN~\citep{zhu2017unpaired}, I-SECRET~\citep{cheng2021secret}, and StillGAN~\citep{ma2021structure} were proposed to learn enhancement models from unpaired data.
{
Synthesized data with segmentation masks were used to present the enhancement algorithm of \cite{luo2020dehaze}, CofeNet~\citep{shen2020modeling}, and MAGE-Net~\citep{guo2023bridging}, whose segmentation modules were learned from DRIVE.
Additionally, RFormer~\citep{deng2022rformer} trained on high-low quality paired real data was also employed in the experiment.
Access to test data was available in the training of I-SECRET~\citep{cheng2021secret}, \cite{luo2020dehaze}, MAGE-Net~\citep{guo2023bridging}, and ArcNet~\citep{li2022annotation}, in which domain adaptation was introduced to bridge the enhancement models from synthesized data to real ones. 
}

\subsection{Enhancement comparisons with SOTA algorithms}
This section presents multiple comparisons with the SOTA algorithms to demonstrate the clinical value of GFE-Net.
Data dependency, enhancement performance, deployment efficiency, and scale generalizability were respectively compared to comprehensively understand the superiority of GFE-Net.

\begin{figure*}[tb]
\centering
\includegraphics[width=\linewidth]{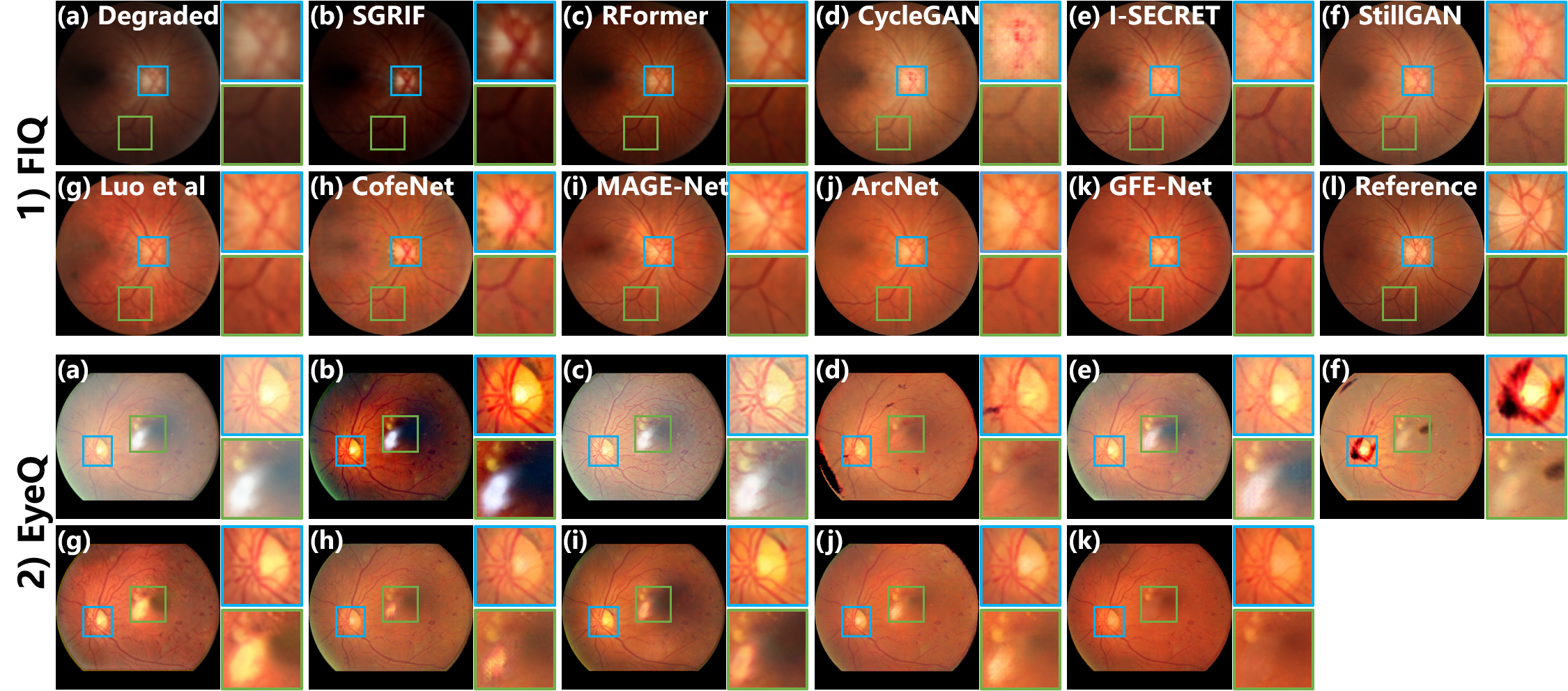}
\caption{{
Fundus image enhancement for imaging interferences. 
A high-quality reference is provided for FIQ.
}}
\label{fig:degradation1}
\end{figure*}

\subsubsection{Data dependency}
Conveniently deploying fundus image enhancement in clinics is a fundamental motivation of this study.
Given that the deployment bottleneck concentrates on the collection of training data, the data dependency of algorithms is compared and summarized in Columns 2-5 of Table~\ref{tab:comparisons}.

As a hand-crafted algorithm, SGRIF~\citep{cheng2018structure} can be directly applied to any fundus image without training, but resulting in limited universality.
{
High-low quality paired fundus images were collected to train RFormer~\citep{deng2022rformer}, and the trained model was directly employed in the experiment.
CycleGAN~\citep{zhu2017unpaired}, I-SECRET~\citep{cheng2021secret}, and StillGAN~\citep{ma2021structure} employ unpaired high-low quality data to learn the enhancement models. In addition, I-SECRET~\citep{cheng2021secret} uses test data during training to boost the generalization. 
Training with synthesized data, \cite{luo2020dehaze}, CofeNet~\citep{shen2020modeling}, MAGE-Net~\citep{guo2023bridging}, ArcNet~\citep{li2022annotation}, and GFE-Net circumvent the dependency on high-low quality data.
However, segmentation models are necessary for ~\cite{luo2020dehaze} and CofeNet~\citep{shen2020modeling}, which utilize the segmentation results to preserve fundus structures.
Test data are accessed during training by \cite{luo2020dehaze}, MAGE-Net~\citep{guo2023bridging}, and ArcNet~\citep{li2022annotation}, to respectively generate synthesized data and adapt real data.
}

Therefore, compared with the SOTA algorithms, GFE-Net minimizes the dependency on training data, thereby boosting clinical deployment of fundus image enhancement.
Furthermore, by eliminating the reliance on on knowledge from test data, the enhancement model learned by GFE-Net can be generically applied to any fundus images without extra tuning.

\begin{figure*}[tbp]
\centering
\includegraphics[width=\linewidth]{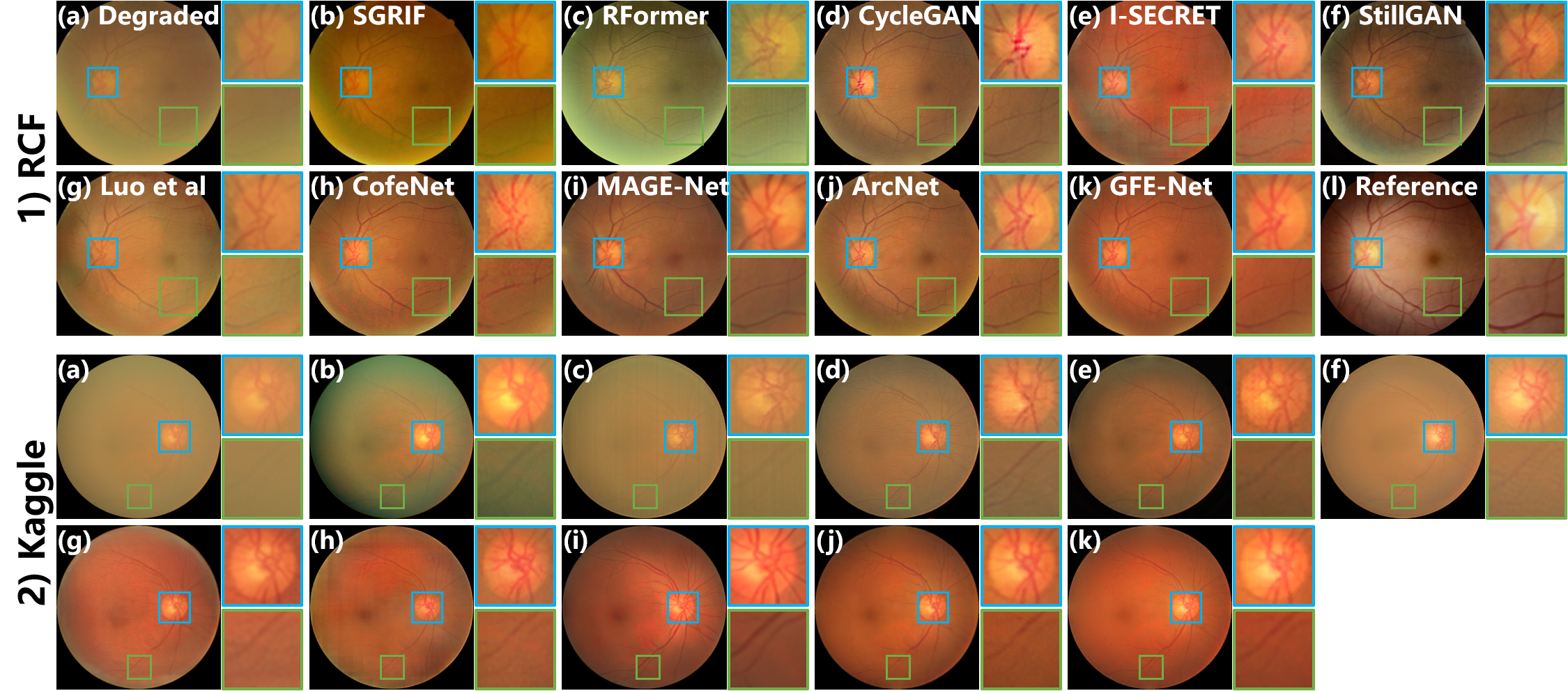}
\caption{{
Fundus image enhancement for cataracts. 
A high-quality reference is provided for RCF.
}}
\label{fig:degradation2}
\end{figure*}

\subsubsection{Enhancement performance}
The enhancement performance was evaluated on datasets with imaging interferences or cataracts, where FIQ and EyeQ suffer from imaging interferences, while RCF and Kaggle are degraded by cataracts.
As exhibited in Table~\ref{tab:comparisons} and Figures~\ref{fig:degradation1} and ~\ref{fig:degradation2}, the performances on FIQ and RCF were evaluated with references, and non-reference evaluations were quantified on EyeQ and Kaggle.

It should be noted that due to the dependency on extra data, the compared SOTA algorithms, except SGRIF~\citep{cheng2018structure} and CofeNet~\citep{shen2020modeling}, were respectively trained with each dataset.
In contrast, GFE-Net was only learned once from synthesized data and generically applied to all datasets.

\noindent \textbf{Imaging-interference degraded data:}
Quantitative comparisons of enhancing FIQ and EyeQ are summarized in Columns 6-9 of Table~\ref{tab:comparisons}, and Fig.~\ref{fig:degradation1} visually compares the correction of imaging interferences.

Hand-crafted spatial filters allow SGRIF~\citep{cheng2018structure} to be efficiently implemented but limit the universality. Thence, distorted color and luminance are observed in Fig.~\ref{fig:degradation1} (b), corresponding to the inferior results in Table~\ref{tab:comparisons}.
{
RFormer~\citep{deng2022rformer} was trained by the paired data collected from clinics, but its generalizability to unseen data is shown to be mediocre.}
Training with unpaired high-low quality data mitigates the data dependency of CycleGAN~\citep{zhu2017unpaired}, I-SECRET~\citep{cheng2021secret}, and StillGAN~\citep{ma2021structure}. 
However, from unpaired data, structure preservation cannot be favorably learned by CycleGAN as shown in Fig.~\ref{fig:degradation1} (d). 
Importance maps and structure loss are hence introduced in I-SECRET~\citep{cheng2021secret} and StillGAN~\citep{ma2021structure} to preserve retinal structures.
Nevertheless, undesired artifacts still appear in the result of StillGAN~\citep{ma2021structure} (Fig.~\ref{fig:degradation1} (f2)).
Synthesized corresponding data are generated to train \cite{luo2020dehaze}, CofeNet~\citep{shen2020modeling}, MAGE-Net~\citep{guo2023bridging}, and ArcNet~\citep{li2022annotation}. Furthermore, structure segmentations are employed in \cite{luo2020dehaze}, CofeNet~\citep{shen2020modeling}, and MAGE-Net~\citep{guo2023bridging} to preserve retinal structures.
Unfortunately, as visualized in Fig.~\ref{fig:degradation1}, segmentation errors resulting from low image quality lead to artifacts in enhanced images.
Decent performance is achieved by MAGE-Net~\citep{guo2023bridging} and ArcNet~\citep{li2022annotation}, which import test data during training to adapt the model from synthesized to real degraded images.

\noindent \textbf{Cataract degraded data:}
Restoring fundus photographs taken through cataracts is clinically valuable to improve the certainty of fundus assessment. 
Columns 10-13 in Table~\ref{tab:comparisons} and Fig.~\ref{fig:degradation2} summarize the restoration performance on RCF and Kaggle.

As collecting high-quality fundus images is impracticable for cataract patients, restoring cataract images without any annotation poses a fundamental challenge for enhancement algorithms. 
Compared to imaging interferences, the severe occlusion caused by cataracts exacerbates the difficulties in image enhancement.
The distorted appearance restored by SGRIF~\citep{cheng2018structure} impacts the evaluation and fundus assessment.
RFormer~\citep{deng2022rformer} learned from paired data cannot be effectively generalized to cataract data.
Vessel loss and fractures are observed in the results from CycleGAN~\citep{zhu2017unpaired}.
Irregular textures commonly appear in the images restored by I-SECRET~\citep{cheng2021secret}, StillGAN~\citep{ma2021structure}, \cite{luo2020dehaze}, and CofeNet~\citep{shen2020modeling}.
Though reasonable performances are presented by MAGE-Net~\citep{guo2023bridging} and ArcNet~\citep{li2022annotation}, visiting test data during training limits the clinical value and deployment efficiency.

By training only once with synthesized data, the proposed GFE-Net demonstrates outstanding performance in both enhancement scenarios.
GFE-Net outperforms the SOTA algorithms as shown in Table~\ref{tab:comparisons} and preserves authentic and clean fundus details in Figures~\ref{fig:degradation1}(k) and~\ref{fig:degradation2}(k).
Specifically, the SOTA algorithms are incapable of properly correcting Fig.~\ref{fig:degradation1}-2), as a result of assembling various types of interference.
Impressively, GFE-Net appropriately processes the degradation, where sparkling artifacts are removed and authentic lesions are preserved.
Moreover, in the evaluations on RCF and Kaggle, GFE-Net even outperforms ArcNet~\citep{li2022annotation}, which is designed to restore cataract images by visiting test data during training. 
In summary, according to Table~\ref{tab:comparisons} and Figures~\ref{fig:degradation1} and ~\ref{fig:degradation2}, the robust representations learned by GFE-Net enable effective enhancement of fundus images degraded by various interferences, without annotations or access to test data.

\subsubsection{Deployment efficiency}
\begin{figure}[tbp]
\centering
\includegraphics[width=0.8\linewidth]{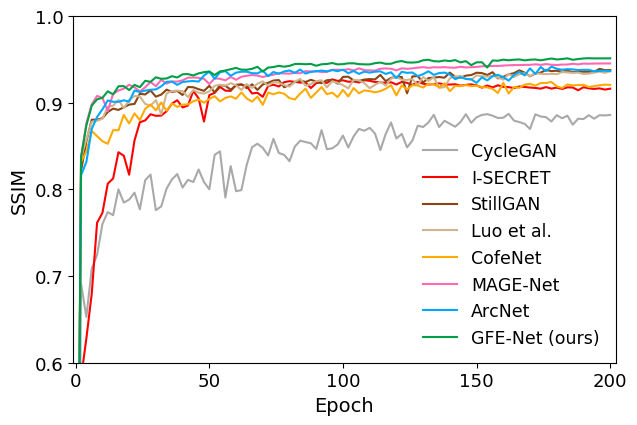}
\caption{
Training history quantified by SSIM on validation data.}
\label{fig:convergence}
\end{figure}

Convenient deployment in clinics is a fundamental precondition for fundus image enhancement algorithms.
The training history is quantified on validation data by SSIM in Fig.~\ref{fig:convergence} to visualize the training efficiency.
SGRIF~\citep{cheng2018structure} and RFormer~\citep{deng2022rformer} are not included in Fig.~\ref{fig:convergence}, since they were not trained in the experiment.
As illustrated in Fig.~\ref{fig:convergence}, GFE-Net is more efficient than SOTA algorithms in achieving convergence.
The SOTA algorithms have extensively conducted adversarial learning to improve enhancement performance.
However, adversarial learning is liable to mode collapse and vanishing gradients, resulting in a time-consuming training phase.
Alternatively, GFE-Net avoids the bottlenecks of adversarial learning by optimizing with explicit objective functions given in Eq.~\ref{eq:overall}.

Moreover, the implementation costs of each algorithm are exhibited in Columns 14-16 of Table~\ref{tab:comparisons} to show the deployment efficiency of GFE-Net.
Specifically, computational costs and time consumption are respectively quantified by multiply-accumulate operation (GMac) as well as training and inference time.
The efficient convergence promises GFE-Net the benefits in training time, while significant progress has been achieved in data dependency and image enhancement by GFE-Net with negligible costs of computation and inference time.
Further, compared to repeatedly training the SOTA algorithms with various datasets, GFE-Net presents an impressive universality, which was trained only once with synthesized data but achieved remarkable performance.

\subsubsection{Scale generalizability}
As clinical data are acquired with various resolutions, the generalizability to different scales is meaningful in clinics.
Thus the enhancement models learned from the image size of 256 $\times$ 256 were further applied to other scales for interpreting the scale generalizability.

\begin{figure}[t]
\centering
\includegraphics[width=\linewidth]{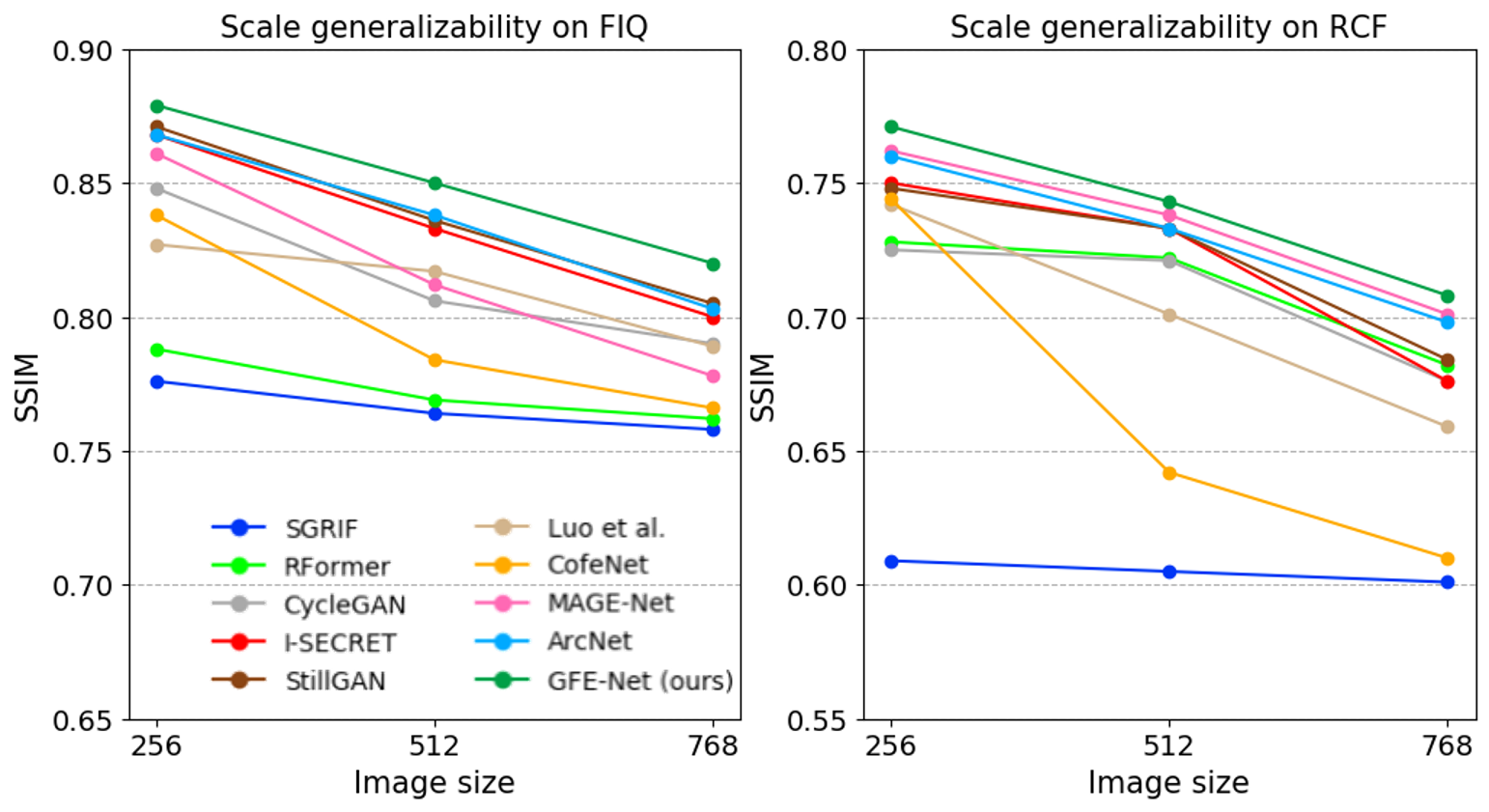}
\caption{
Scale generalizability of the enhancement models. The models trained with the scale of 256 $\times$ 256 are also applied on the scales of 512 $\times$ 512 and 768 $\times$ 768.
}
\label{fig:scale}
\end{figure}

The models were applied to the scales of 512 $\times$ 512 and 768 $\times$ 768, and the SSIM of FIQ and RCF was employed to demonstrate the scale generalizability.
Fig.~\ref{fig:scale} shows the performance on various image scales. 
Compared with training and testing with identical scales, applying the models to scales larger than training images suppresses the enhancement performance.
Fortunately, outstanding scale generalizability is provided by the proposed algorithm, which outperforms the SOTA algorithms on each scale.
Accordingly, GFE-Net is more convenient to adapt various scales of test data.

\subsection{Boosting for fundus image analysis}
Besides improving fundus observation, another primary goal of fundus image enhancement is to facilitate clinical examination and diagnosis. 
Therefore, automatic analyses were carried out on the enhanced images to explore the clinical value of the proposed algorithm.
Specifically, retinal vessel segmentation was conducted on FIQ and RCF, and fundus disease diagnosis was performed on Fundus-iSee.

\subsubsection{Retinal vessel segmentation}
\begin{figure*}[bt]
\centering
\includegraphics[width=\linewidth]{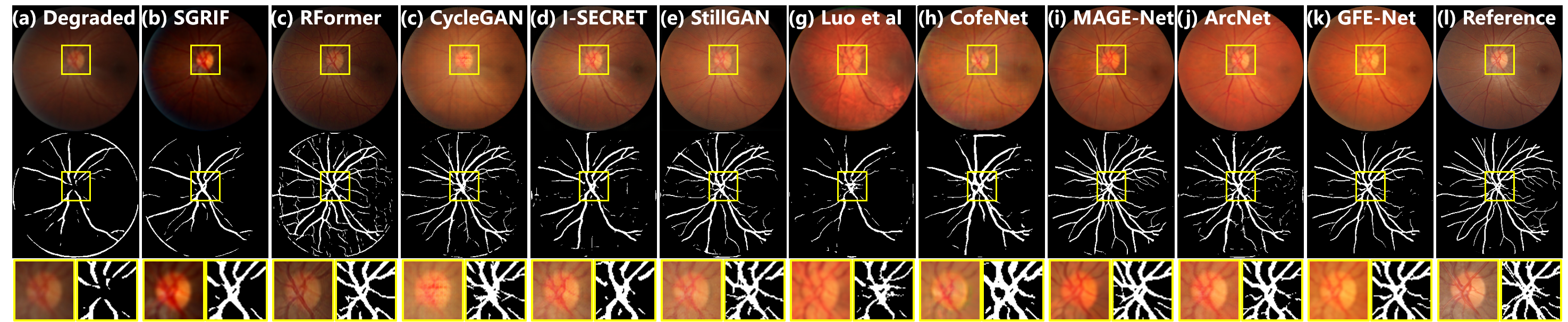}
\caption{{
Segmentation comparison of enhanced fundus images. 
The reference segmentation (j) is used as the ground truth for quantitative metrics.
}
}
\label{fig:seg}
\end{figure*}

A segmentation model was trained on DRIVE using U-Net and then applied to segment the fundus images in FIQ and RCF.
To evaluate the segmentation performance on enhanced images, the segmentation metrics of IoU and Dice were calculated regarding the segmentation results of the high-quality references in FIQ and RCF as ground truth.
The segmentation results of low-quality images were also provided as a benchmark.
Table~\ref{tab:seg} summarizes the quantitative results, and Fig.~\ref{fig:seg} exhibits the visual ones.

\begin{table}[b]
\footnotesize
\centering {\caption{{Boosting Segmentation via Image Enhancement.}}
\label{tab:seg} }%
\renewcommand{\arraystretch}{1.1}
\begin{tabular}{p{37mm} p{6mm}<{\centering} p{6mm}<{\centering} p{0.6mm} p{6mm}<{\centering} p{6mm}<{\centering}}
\hline
\multirow{2}{*}{Algorithms} & \multicolumn{2}{c}{FIQ} && \multicolumn{2}{c}{RCF}\\
\cline{2-3}
\cline{5-6}
& IoU & Dice && IoU & Dice\\
\hline
Low-quality & 0.179 & 0.304 && 0.350 & 0.518\\
\hline
\hline
SGRIF~\citep{cheng2018structure}& 0.410 & 0.580 && 0.194 & 0.402\\
RFormer~\citep{guo2023bridging}& 0.371 & 0.542 && 0.208 & 0.344\\
CycleGAN~\citep{zhu2017unpaired}& 0.352 & 0.520 && 0.256 & 0.408\\
\cite{luo2020dehaze}& 0.335 & 0.502 && 0.306  & 0.469\\
CofeNet~\citep{shen2020modeling}& 0.436 & 0.607 &&  0.360 & 0.529\\
I-SECRET~\citep{cheng2021secret}& 0.453 & 0.623 && 0.371 & 0.541\\
StillGAN~\citep{ma2021structure} & 0.450 & 0.620 && 0.373 & 0.544\\
MAGE-Net~\citep{guo2023bridging}& \textbf{0.472} & \textbf{0.646} && 0.414 & 0.585\\ 
ArcNet~\citep{li2022annotation} & 0.456 & 0.627 && 0.401 & 0.572\\
GFE-Net (ours) & 0.468 & 0.638 && \textbf{0.418} & \textbf{0.587}\\
\hline
\end{tabular}
\end{table}

According to Table~\ref{tab:seg} and Fig.~\ref{fig:seg}, most enhancement algorithms successfully improve the segmentation performance compared to low-quality images, which verifies the effectiveness of image restoration for automatic fundus analysis.
{
Notably, it is reasonable that MAGE-Net~\citep{guo2023bridging} provides remarkable performance on FIQ, as it loads both segmentation masks and test data during training.
}
Due to the favorable structure preservation in GFE-Net, a superior segmentation performance to SOTA algorithms is achieved in the results of GFE-Net.
Accordingly, the proposed algorithm is capable of facilitating automatic fundus analysis and examination.

\subsubsection{Fundus disease diagnosis}
The automatic diagnosis was implemented on Fundus-iSee to assess the value of enhancement for fundus examination.
Images in Fundus-iSee are annotated into five categories: AMD, DR, glaucoma, high myopia, and normal fundus.
Five thousand clear images were randomly split from Fundus-iSee to train the enhancement and diagnosis models, and the rest clear and cataract images were respectively used for testing. 
Subsequently, cataract images were enhanced by the algorithms to compare the promotion for fundus disease diagnosis.
F1-score and CKappa were used to quantify the diagnosis performance on the clear, cataract, and enhanced images, as summarized in Table~\ref{tab:isee}. 

Due to the distorted fundus observation caused by cataracts, 
the risk of misdiagnosis is increased in cataract images compared to clear images, leading to a significant decrease in diagnosis accuracy. 
\begin{table}[tb]
\footnotesize
\centering {\caption{{Boosting Diagnosis via Image Enhancement.}}
\label{tab:isee} }%
\renewcommand{\arraystretch}{1.1}
\begin{tabular}{p{40mm} p{15mm}<{\centering} p{15mm}<{\centering}}
\hline
Algorithms & F1-score & Ckappa\\
\hline
Clear & 0.838 & 0.448\\
Cataract & 0.730 & 0.310\\
\hline
\hline
SGRIF~\citep{cheng2018structure} & 0.760 & 0.420\\
RFormer~\citep{deng2022rformer} & 0.732 & 0.370 \\
CycleGAN~\citep{zhu2017unpaired} & 0.724 & 0.286\\
\cite{luo2020dehaze} & 0.712 & 0.370\\
CofeNet~\citep{shen2020modeling} & 0.754 & 0.416\\
I-SECRET~\citep{cheng2021secret} & 0.734 & 0.382 \\
StillGAN~\citep{ma2021structure} & 0.739 & 0.348 \\
MAGE-Net~\citep{guo2023bridging} & 0.753 & 0.390 \\
ArcNet~\citep{li2022annotation} & 0.761 & 0.428\\
GFE-Net (ours)& \textbf{0.771} & \textbf{0.445}\\
\hline
\end{tabular}
\end{table}
The performances of clear and cataract images in Table~\ref{tab:isee} separately sketch the upper and lower benchmarks of diagnosis on Fundus-iSee.
By improving the readability of cataract images, the fundus disease diagnosis is boosted by most enhancement algorithms. 
Despite the distinctive color and illuminance of the results from SGRIF~\citep{cheng2018structure}, fundus lesions are reasonably enhanced for the diagnosis model.
Desirable performance is achieved on the images enhanced by GFE-Net, indicating that the proposed modules properly preserve and enhance fundus lesions.

\subsection{Ablation study}
Ablation studies on training data and designed modules were conducted to demonstrate the impact and effectiveness of the modules.

\subsubsection{Training data}
\begin{figure}[tb]
\centering
\includegraphics[width=0.95\linewidth]{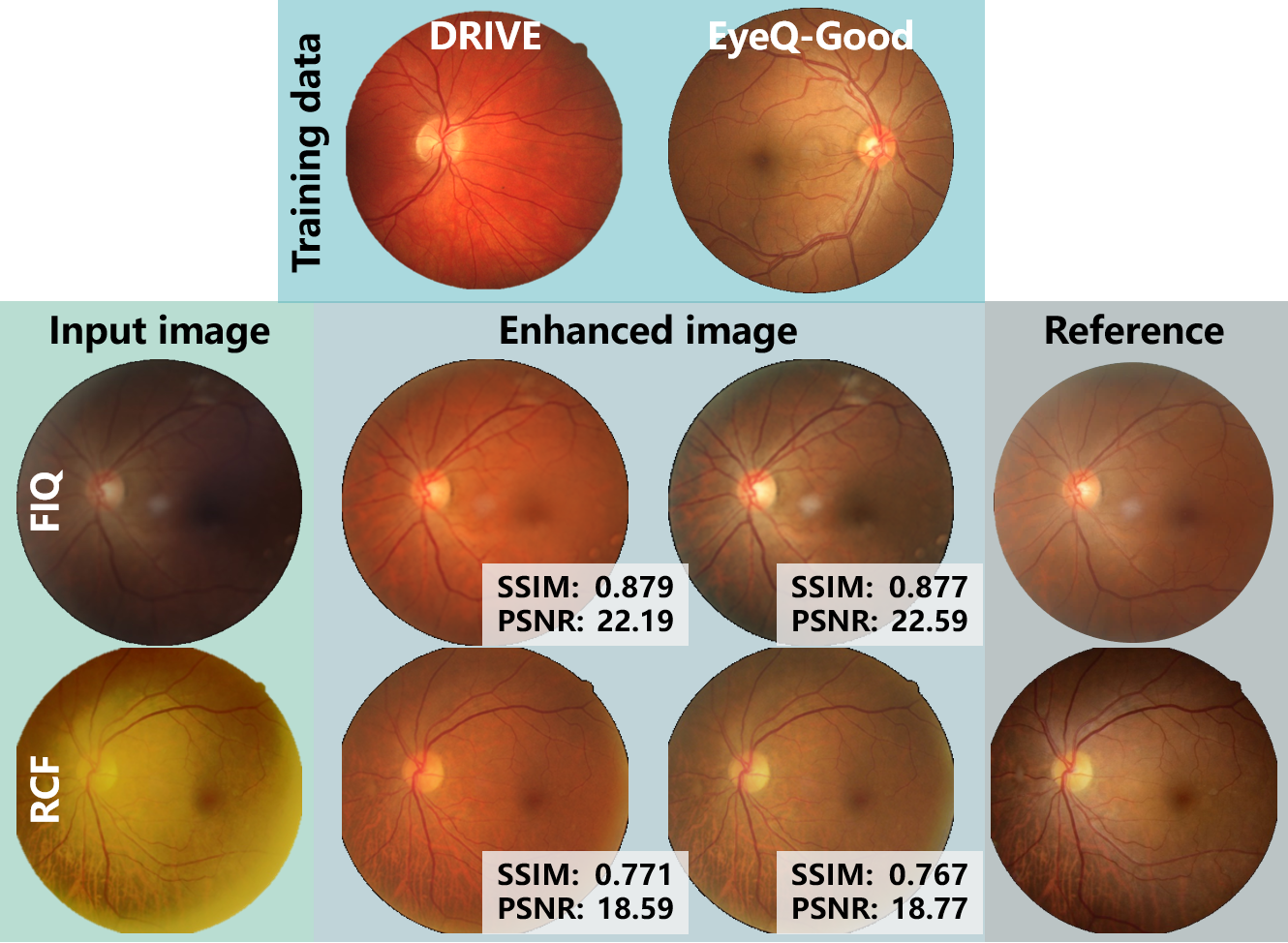}
\caption{
Comparison of the enhancement models learned from different training data. Quantitative performances are also provided by SSIM and PSNR.}
\label{fig:source}
\end{figure}

As GFE-Net is trained without supervised or extra data, the enhanced images are only regulated by the clear images, which are used to synthesize training data.
To understand the impact of the training data, different clear image datasets were employed to synthesize high-quality image pairs for training.
Fig.~\ref{fig:source} visualizes the variant outcomes resulting from the training data synthesized from DRIVE and EyeQ, and a quantitative comparison is also provided.

Images are properly enhanced by both models learned from training data synthesized from DRIVE and EyeQ, and comparable quantitative performances are achieved. 
Thence, GFE-Net is insensitive to the training data selection, since robust structure-aware representations are learned by SSRL.
However, as exhibited in Fig.~\ref{fig:source}, style discrepancy is observed from the enhanced images.
Though SSRL captures structure-aware representations in GFE-Net, the final enhancement results are supervised by the clear datasets used to generate training data.
Thus it is no surprise that the image style of enhanced images follows the clear ones.
{
Notably, the structures in fundus images play a more critical role in diagnosis than the image styles. Although various styles in the training data may affect the styles of enhancement outcomes, their impact on the downstream analysis is minimal.
}


\subsubsection{Designed modules}
\begin{figure}[b]
\centering
\includegraphics[width=0.95\linewidth]{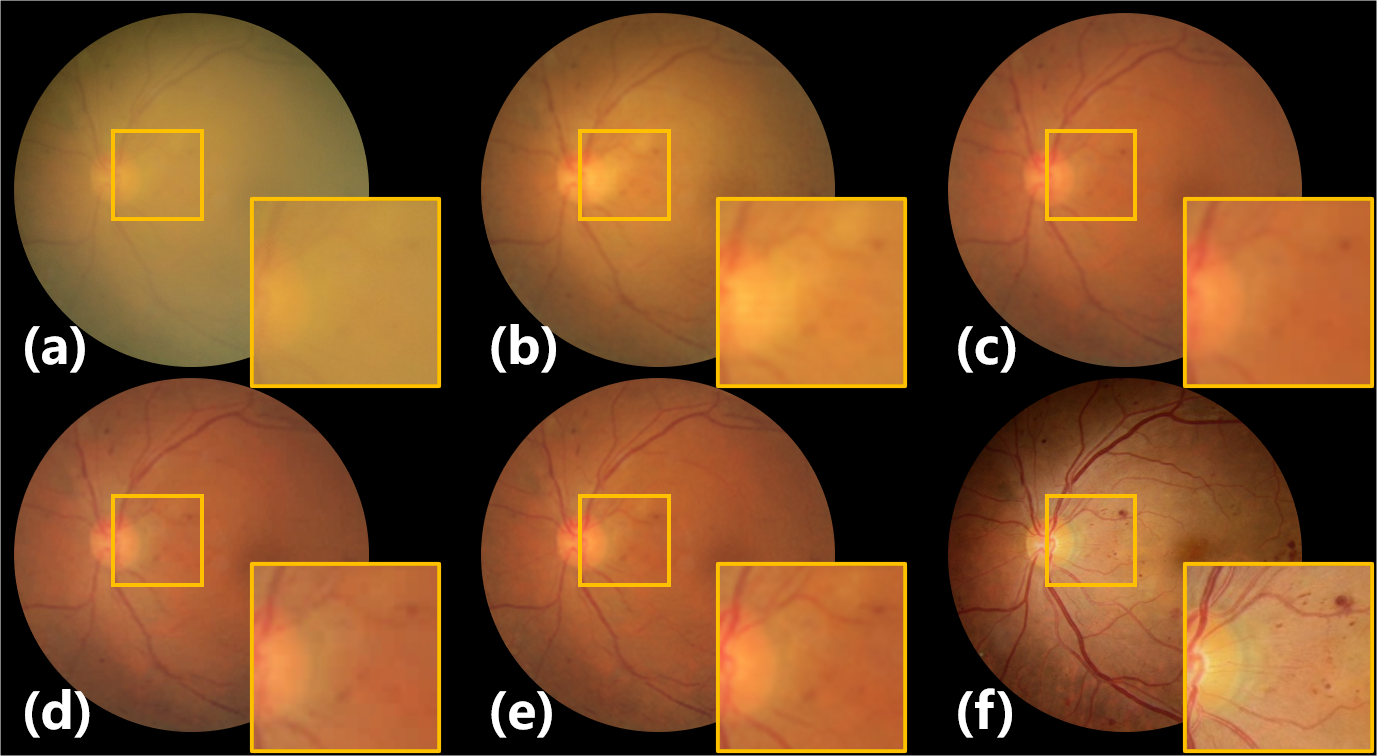}
\caption{
The effect of the proposed modules. (a) cataract image. (b) w/o $\mathcal{F}(\cdot)$, $\mathcal{L}_R$ and $\mathcal{L}_{cyc}$. (c) w/o $\mathcal{L}_R$ and $\mathcal{L}_{cyc}$. (d) w/o $\mathcal{L}_{cyc}$. (e) GFE-Net. (f) reference.}
\label{fig:ablation}
\end{figure}

Comparisons against the ablation of modules were conducted to validate the effectiveness of the designed modules.
The modules were successively installed according to the dependencies between them.
In the beginning, removing all the proposed modules from GFE-Net, a 
U-Net was introduced as the backbone for fundus image enhancement.
Then three modules were installed step by step to promote the enhancement, including (1) $\mathcal{F}(\cdot)$ to extract frequency information for highlighting retinal structures,
(2) $\mathcal{L}_R$ to learn structure-aware representations under the supervision from $\mathcal{F}(\cdot)$ to robustly correct fundus images, and
(3) $\mathcal{L}_{cyc}$ to impose cycle-consistency between the outcomes of $D_R$ and $D_E$ to further facilitate the network optimization.
Fig.~\ref{fig:ablation} visualizes the images enhanced by various module combinations, and quantitative evaluations on FIQ and RCF are summarized in Table~\ref{tab:Ablation}.
Due to the discrepancy between DRIVE and RCF, appearance gap is inevitable between the reference (Fig.~\ref{fig:ablation} (f)) and the enhanced images (Fig.~\ref{fig:ablation} (b-e)). 
Therefore, we concentrate on analyzing the readability of enhanced images, as the appearance gap has little impact on fundus assessment.

\begin{table}[htb]
\footnotesize
\centering
\caption{Ablation study of the proposed modules.}
\label{tab:Ablation}
\renewcommand{\arraystretch}{1.1}
\begin{tabular}{p{0.6cm}<{\centering} p{0.6cm}<{\centering} p{0.6cm}<{\centering}|| p{0.6cm}<{\centering} p{0.6cm}<{\centering} p{0.6mm}<{\centering} p{0.6cm}<{\centering} p{0.6cm}<{\centering}}
\hline
\multirow{2}{*}{$\mathcal{F}(\cdot)$} & \multirow{2}{*}{$\mathcal{L}_R$} & \multirow{2}{*}{$\mathcal{L}_{cyc}$} & \multicolumn{2}{c}{FIQ} && \multicolumn{2}{c}{RCF}\\
\cline{4-5}
\cline{7-8}
& & & SSIM & PSNR && SSIM & PSNR\\
\hline
& & & 0.845 & 20.59 && 0.730 & 17.73 \\
$\surd$ &  & & 0.859 & 21.27 && 0.752 & 18.17 \\
$\surd$ & $\surd$ & & 0.866 & 21.87 && 0.765 & 18.39 \\
$\surd$ & $\surd$ & $\surd$ & \textbf{0.879} & \textbf{22.19} && \textbf{0.771} & \textbf{18.59} \\
\hline
\end{tabular}
\end{table}

Compared to the reference in Fig.~\ref{fig:ablation} (f), the quality of Fig.~\ref{fig:ablation} (a) has been severely degraded by cataracts.
Using the backbone of U-Net, image contrast has been increased in Fig.~\ref{fig:ablation} (b), but fundus structures are still too fuzzy to observe.
Thanks to the frequency information from $\mathcal{F}(\cdot)$, the readability of fundus structures has been improved in Fig.~\ref{fig:ablation} (c), though fine retinal vessels and lesions are neglected in the zoomed box.
Then $\mathcal{L}_R$ executes SSRL under frequency supervision, forwarding robust structure-aware representations to the network.
Retinal vessels and hemorrhage lesions are thus visible in the zoomed box of Fig.~\ref{fig:ablation} (d).
Finally, $\mathcal{L}_{cyc}$ further facilitates the network optimization such that fundus structures and pathologies have been decently enhanced in Fig.~\ref{fig:ablation} (e).
Correspondingly, in Table~\ref{tab:Ablation}, the enhancement performance progresses with module implementation, further confirming the effectiveness of the proposed modules.

\subsection{Discussion}
Through the experiments, GFE-Net is demonstrated to be an efficient generic algorithm for fundus image enhancement, which robustly enhances unknown fundus images and preserves retinal structures without supervised or extra data.

Comprehensive comparisons, including data dependency, enhancement performance, deployment efficiency, and scale generalizability, were conducted against SOTA algorithms to verify the advantages of GFE-Net in clinical deployment.
{
Compared with SOTA algorithms, our method effectively reduces data dependency by eliminating the need for hand-crafted features, segmentation masks, unpaired training data, or access to test data in the deployment of GFE-Net. 
Additionally, GFE-Net achieves superior enhancement performance with reasonable computing costs. 
By significantly improving the readability of fundus images, GFE-Net greatly enhances the subsequent analysis of these images.
Furthermore, the ablation study validates the effectiveness of the designed modules in GFE-Net and highlights the impact of training data.
}

In contrast to the various data dependency of SOTA algorithms, only synthesized data are necessary for GFE-Net.
Furthermore, the data dependency of SOTA algorithms often necessitates repeated training, whereas the model from GFE-Net can be directly applied to any fundus images without the need for additional tuning.
Specifically, compared with the previous study~\citep{li2022annotation}, SSRL allows GFE-Net to circumvent the dependency on test data, such that desirable enhancement models are learned only based on synthesized data.
Moreover, the seamless SSRL architecture improves the robustness and efficiency of task-relevant models, where GFE-Net outperforms ArcNet~\citep{li2022annotation} in enhancement performance, deployment efficiency, and scale generalizability.
However, the image style of outcomes in GFE-Net deeply depends on the training data.
Although the image style has little impact on fundus observation, automatic fundus analysis suffers from domain shifts resulting from image styles.

\section{Conclusion}
Image degradation impacts fundus observation, leading to uncertainty in fundus assessment. 
Although algorithms have been developed to enhance fundus images, cumbersome data dependency and inconvenient applicability limit the clinical deployment of fundus enhancement.
To address this bottleneck, a generic enhancement network, named GFE-Net, was proposed in this study to robustly correct unknown fundus images without dependency on supervised or extra data.
By designing a seamless SSRL architecture, GFE-Net couples frequency self-supervision-based representation learning with down-steam image enhancement.
Hence robust structure-aware representations were learned only based on synthesized data to outstandingly enhance fundus images suffering from unknown degradations.
Comprehensive experiments showed the effectiveness and advantages of GFE-Net. 
GFE-Net not only outperformed the SOTA algorithms in correcting fundus images, but also boosted clinical deployment and follow-up fundus image analysis.

\section*{Acknowledgments}
This work was supported in part by Guangdong Basic and Applied Basic Research Foundation (2020A1515110286), National Natural Science Foundation of China (82102189), Guangdong Provincial Department of Education (2020ZDZX3043), Guangdong Provincial Key Laboratory (2020B121201001), and Shenzhen Natural Science Fund (JCYJ20200109140820699, 20200925174052004).

\bibliographystyle{model2-names.bst}\biboptions{authoryear}
\bibliography{refs}



\end{document}